\documentclass[aps,prl,showpacs,twocolumn,superscriptaddress]{revtex4}

\usepackage{amsmath}
\usepackage{amssymb}
\usepackage{bm}
\usepackage{natbib}
\usepackage{graphicx}
\usepackage{color}
\usepackage{dcolumn}

\begin{document}


\title{Design of Low Band Gap Double Perovskites from First Principles}

\author{Robert F. Berger}
\affiliation{Molecular Foundry, Lawrence Berkeley National Laboratory, Berkeley, CA}
\author{Jeffrey B. Neaton}
\email{jbneaton@lbl.gov}
\affiliation{Molecular Foundry, Lawrence Berkeley National Laboratory, Berkeley, CA}


\begin{abstract}
Using density functional theory (DFT)-based calculations, we propose a family of metastable, as-yet unmade V$^{5+}$ and Cr$^{6+}$ double perovskite compounds with low band gaps spanning much of the visible region of the solar spectrum. Through analysis of a related set of measured optical gaps of $d^{0}$ ABO$_{3}$ perovskites and A$_{2}$B'BO$_{6}$ double perovskites, an {\em ad hoc} procedure is developed to correct DFT and many-body perturbation theory gaps, bringing them into quantitative agreement with experiment for measured compounds, and predicting that V$^{5+}$ and Cr$^{6+}$ double perovskites would have gaps ranging from approximately 1.1--2.4 eV, significantly lower than previous materials studied in this class. DFT calculations also establish that these V$^{5+}$ and Cr$^{6+}$ compounds are likely able to be synthesized, either in bulk form or as epitaxial thin films. These compounds would comprise a new class of semiconducting double perovskites for potential use in solar energy conversion and other optoelectronic applications.
\end{abstract}

\date{\today}

\pacs{}

\maketitle


\section{Introduction}

The ability to tune the electronic structure of perovskite oxides is of significant interest in areas ranging from superconductivity to thermoelectricity to solar energy conversion. For example, perovskite-derived transition metal oxide compounds have been shown to photocatalyze the splitting of water into hydrogen and oxygen gas.~\cite{kudo09csr,yerga09csc} However, their ability to harvest sunlight in the visible spectrum is primarily limited by their large optical band gaps. Experimental efforts to reduce perovskite optical gaps have included chemical doping and substitution.~\cite{fujimori92prb,fujimori92jpcs,kato02jpcb,kasahara02jpca,yin03jpcb,konta04jpcb,hur05jpcb,kim07mrb} Especially relevant to this work, Eng {\em et al.}~\cite{eng03jssc} synthesized a number of $d^{0}$ perovskites (ABO$_{3}$) and double perovskites (A$_{2}$B'BO$_{6}$) with rock-salt B-site ordering. Still, none of these compounds possess gaps below 2.7 eV, leaving open the question of whether there might exist materials in this class with more significant overlap with the solar spectrum.

First-principles calculations based on density functional theory (DFT) have proven useful for exploring the large structural parameter space of chemically doped and substituted perovskites to search for new materials with reduced band gaps.~\cite{fujimori92prb,fujimori92jpcs,bennett08jacs,bennett09prb,qi11prb,castelli12ees} However, since DFT is well known to underestimate band gaps, the quantitative nature of these predictions is limited. Many-body perturbation theory within the $GW$ approximation can result in significantly more reliable gaps for many inorganic semiconductors,~\cite{hybertsen86prb,shishkin07prb} despite its neglect of excitonic effects and electron-phonon interactions. Yet standard implementations of the $GW$ approximation---in particular one-shot $G_{0}W_{0}$ corrections---have been less successful for transition metal oxides where $d$ states feature prominently.~\cite{tiago04prb,vanschilfgaarde06prl,shishkin07prb,kang10prb,shih10prl,friedrich11prb,berger11prl,stankovski11prb,patrick12jpcm} Thus, quantitative first-principles prediction of band gaps for transition metal oxides remains a principal challenge.

In this work, we propose a class of new double perovskite transition metal oxides with band gaps well within the visible region of the solar spectrum, mitigating difficulties in using conventional DFT-based methods to quantitatively assess transition metal oxide band gaps with an empirical approach. We first identify a group of as-yet unmade $d^{0}$ double perovskites containing V$^{5+}$ or Cr$^{6+}$ at the B sites that are metastable with respect to structural distortions and changes in B-site ordering. Building on existing optical data on gaps of 30 different $d^{0}$ perovskites and double perovskites,~\cite{eng03jssc} we then develop and use an {\em ad hoc} procedure, based on a fit to the difference between DFT and experimental gaps that corrects for DFT errors in electron correlation and exchange, to predict optical gaps to within about $\pm0.2$ eV. These {\em ad hoc} corrections are used to predict optical gaps of metastable V$^{5+}$ and Cr$^{6+}$ compounds in the middle or toward the lower end of the visible region of the solar spectrum, establishing these new double perovskites' potential for solar energy conversion and other optoelectronic applications.

\section{Computational Details}

All crystal structures and total energies are computed using DFT within the local density approximation (LDA), the VASP package,~\cite{kresse96prb} and PAW potentials~\cite{kresse99prb}.~\footnote{In VASP calculations throughout this work, electrons taken to be valence are $2s^{2}2p^{4}$ of O, $2p^{6}3s^{2}$ of Mg, $3s^{2}3p^{1}$ of Al, $3s^{2}3p^{6}4s^{1}$ of K, $3s^{2}3p^{6}4s^{2}$ of Ca, $3s^{2}3p^{6}4s^{2}3d^{1}$ of Sc, $3s^{2}3p^{6}4s^{2}3d^{2}$ of Ti, $3s^{2}3p^{6}4s^{2}3d^{3}$ of V, $3p^{6}4s^{2}3d^{4}$ of Cr, $4s^{2}3d^{10}$ of Zn, $4s^{2}3d^{10}4p^{1}$ of Ga, $4s^{2}4p^{6}5s^{2}$ of Sr, $4s^{2}4p^{6}5s^{2}4d^{1}$ of Y, $5s^{2}4d^{3}$ of Nb, $5s^{2}4d^{4}$ of Mo, $5s^{2}5p^{6}6s^{2}$ of Ba, $5s^{2}5p^{6}6s^{2}5d^{1}$ of La, $5p^{6}6s^{2}5d^{3}$ of Ta, and $5p^{6}6s^{2}5d^{4}$ of W.} A plane-wave cutoff of 500 eV is used throughout. Calculations on 5-atom perovskite unit cells use a $\Gamma$-centered $10\times10\times10$ $k$-point mesh, while larger unit cells use correspondingly fewer $k$-points. Zone-center phonon frequencies are computed using density functional perturbation theory. Elastic constants and phonon dispersion relations are calculated using a finite differences approach and the Phonopy code.~\cite{togo08prb} $G_{0}W_{0}$ corrections to DFT-LDA eigenvalues (referred to hereafter as $G_{0}W_{0}$@LDA) are evaluated in the LDA optimized crystal structure, using a frequency-dependent dielectric function calculated within the random phase approximation, sampling 100 frequency points; 144 total bands per 5-atom perovskite unit cell (approximately 20 of which are occupied, depending on chemical composition); and a plane-wave cutoff for the response function of 200 eV. As an additional point of comparison, hybrid HSE06~\cite{krukau06jcp} band gaps are calculated using HSE06 optimized crystal structures. Both $G_{0}W_{0}$@LDA and HSE06 calculations on 5-atom perovskite unit cells use a $\Gamma$-centered $6\times6\times6$ $k$-point mesh, while larger unit cells use correspondingly fewer $k$-points.

\section{Results and Discussion}

\subsection{Selection of novel perovskite-derived compounds}

Based on measurements of 31 $d^{0}$ perovskite-derived compounds, Eng {\em et al.}~\cite{eng03jssc} rationalized that their optical band gaps depend primarily on the ``effective electronegativity'' of the B-site transition metal cation. That is, smaller, more highly charged metal cations result in lower-energy conduction bands and correspondingly smaller gaps. We use this principle to infer that V$^{5+}$ and Cr$^{6+}$-based compounds would have {\it even smaller} band gaps than those previously synthesized, allowing for strong absorption in the solar spectrum. Indeed, there exist V$^{5+}$ and Cr$^{6+}$ non-perovskite oxides with optical gaps in the visible region, such as V$_{2}$O$_{5}$ (2.3 eV)~\cite{bodo67pss} and SrCrO$_{4}$ (2.44 eV).~\cite{yin03cpl} The discovery of more materials, especially in the highly tunable perovskite class, could be extremely valuable. It should be noted, however, that toxicity is a possible drawback of the widespread use of V$^{5+}$ and Cr$^{6+}$-based compounds.~\cite{beyersmann08at}

\begin{figure*}
  \begin{center}
    \includegraphics[scale=0.2]{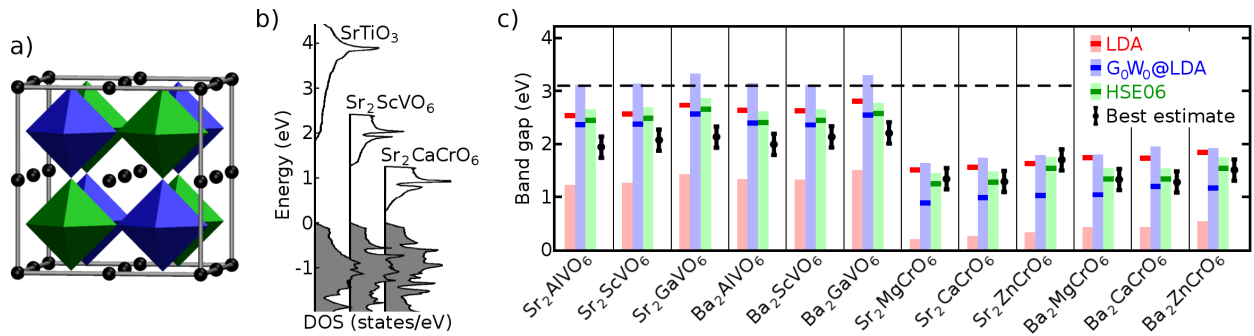}
  \end{center}
  \caption{a) The double perovskite structure, with rock-salt B-site ordering. b) LDA densities of states (DOS) of SrTiO$_{3}$, Sr$_{2}$ScVO$_{6}$, and Sr$_{2}$CaCrO$_{6}$, in which the valence band maxima are defined as zero energy. c) LDA (light red), $G_{0}W_{0}$@LDA (light blue), and HSE06 (light green) band gaps of twelve hypothetical V$^{5+}$ and Cr$^{6+}$ double perovskites with rock-salt B-site ordering. Estimated optical gaps (red, blue, green) based on the assumption that computational errors are equal to those of SrTiO$_{3}$. Estimated optical gaps (black) based on the modification of LDA gaps using linear corrections in $\frac{1}{\epsilon_\infty}$ and $\Delta_\mathrm{X}$ (see text). Error bars ($\pm 0.20$ eV) indicate uncertainty, as determined by the standard deviations of the fits for known compounds. The dashed line represents the optical gap of SrTiO$_{3}$.}
  \label{fig:hypothetical}
\end{figure*}

In what follows, we explore the band gaps of twelve hypothetical V$^{5+}$ and Cr$^{6+}$ double perovskites with rock-salt B-site ordering (Fig.~\ref{fig:hypothetical}a)---(Sr,Ba)$_{2}$(Al,Sc,Ga)VO$_{6}$ and (Sr,Ba)$_{2}$(Mg,Ca,Zn)CrO$_{6}$. From our calculations of phonon frequencies, elastic constants, and cation ordering energies summarized in the Appendix, these twelve hypothetical cubic double perovskites are likely metastable at room temperature. Full analysis of the global stability of these structures would require double perovskite energies to be compared to those of numerous other phases (e.g., binary and ternary oxides). However, their metastability suggests that these compounds could in principle be synthesized, if not in bulk then via epitaxial growth.

In addition to being metastable, the predicted electronic structure of these new V$^{5+}$ and Cr$^{6+}$-based compounds (Fig.~\ref{fig:hypothetical}b) confirms the intuition that smaller, more highly charged metal cations result in lower-energy conduction bands and correspondingly smaller gaps. However, the LDA gaps implied by Fig.~\ref{fig:hypothetical}b are expected to be underestimated and therefore not quantitative, leaving open the question of how much these chemical substitutions might reduce the optical gap relative to SrTiO$_{3}$. A first estimate of their optical gaps (Fig.~\ref{fig:hypothetical}c) is quite promising: assuming that their LDA, $G_{0}W_{0}$@LDA, and HSE06 gaps have the same errors as for SrTiO$_{3}$ relative to room-temperature absorption experiments (i.e., $-1.30$, 0.76, and 0.21 eV for LDA, $G_{0}W_{0}$@LDA, and HSE06, respectively), all twelve compounds are expected to have significantly smaller gaps than SrTiO$_{3}$ that span considerable regions of the solar spectrum, ranging from 2.8 eV down to 0.8 eV in these initial estimates. However, these estimates are severely limited, particularly given that LDA, HSE06, and $GW$ are not rigorous theories of room-temperature optical gaps; additionally, they carry unknown uncertainties. 

\subsection{Corrections of DFT band gaps to experiment}

In the remainder of this article, we develop and apply an {\em ad hoc} correction to DFT and many-body perturbation theory gaps of $d^{0}$ A$_{2}$B'BO$_{6}$ double perovskites, establishing their optical gaps within $\pm 0.20$ eV as significantly lower than previous materials studied in this class. We fit the difference between computed and measured gaps of a set of $d^{0}$ perovskite-derived compounds measured by Eng {\em et al.}~\cite{eng03jssc}, using them as a test set for our {\em ad hoc} correction. The compounds measured by Eng {\em et al.}~\cite{eng03jssc} include a wide range of elements, their electronic structures are qualitatively similar, and their optical gaps have been measured using a consistent procedure. Of the 31 compounds measured, we compute 30 (shown in the top section of Table~\ref{table:testset}) in LDA and $G_{0}W_{0}$@LDA, excluding only NaNbO$_{3}$ because its large 40-atom unit cell~\cite{sakowski69ac} is currently a computational challenge for $G_{0}W_{0}$. Because of the additional computational expense associated with optimizing crystal structures with hybrid functionals, we compute only the twelve cubic compounds with HSE06. Structures are optimized in their experimentally determined room-temperature space groups with rock-salt B-site ordering.

\begin{table*}\scriptsize
  \caption{(Top) Perovskite-derived compounds with $d^{0}$ B-site metal cations considered in this work, with optical gaps taken from the experiments of Eng {\em et al.}~\cite{eng03jssc} (Bottom) Newly-predicted V$^{5+}$ and Cr$^{6+}$ double perovskites. All estimated gaps are based on Eqn.~\ref{eqn:est}, and have uncertainties of $\pm 0.2$ eV.}
  \centering
  \begin{tabular}{l@{\hspace{2ex}}l@{\hspace{2ex}}l@{\hspace{2ex}}l@{\hspace{2ex}}l@{\hspace{8ex}}l@{\hspace{2ex}}l@{\hspace{2ex}}l@{\hspace{2ex}}l@{\hspace{2ex}}l@{\hspace{8ex}}l@{\hspace{2ex}}l@{\hspace{2ex}}l@{\hspace{2ex}}l@{\hspace{2ex}}l}
  \hline
  & & Est. & Opt. & & & & Est. & Opt. & & & & Est. & Opt. & \\
  & Space & gap & gap & & & Space & gap & gap & & & Space & gap & gap & \\
  & group & (eV) & (eV) & Ref. & & group & (eV) & (eV) & Ref. & & group & (eV) & (eV) & Ref.\\
  \hline
  SrTiO$_{3}$ & $Pm\overline{3}m$ & 2.9 & 3.1 & \cite{vousden51ac} & Ba$_{2}$MgMoO$_{6}$ & $Fm\overline{3}m$ & 2.6 & 2.7 & \cite{day12jssc} & Ba$_{2}$CaWO$_{6}$ & $I4/m$ & 3.5 & 3.6 & \cite{day12jssc}\\
  KTaO$_{3}$ & $Pm\overline{3}m$ & 3.3 & 3.5 & \cite{vousden51ac} & Ba$_{2}$CaMoO$_{6}$ & $Fm\overline{3}m$ & 2.6 & 2.7 & \cite{day12jssc} & La$_{2}$MgTiO$_{6}$ & $P2_{1}/n$ & 4.0 & 3.9 & \cite{levin05cm}\\
  KNbO$_{3}$ & $Bmm2$ & 2.7 & 3.1 & \cite{katz67ac} & Ba$_{2}$ZnMoO$_{6}$ & $Fm\overline{3}m$ & 2.9 & 2.7 & \cite{day12jssc} & Ca$_{2}$ScNbO$_{6}$ & $P2_{1}/n$ & 4.1 & 4.1 & \cite{barnes06ac}\\
  CaTiO$_{3}$ & $Pcmn$ & 3.4 & 3.3 & \cite{koopmans83ac} & Ba$_{2}$MgWO$_{6}$ & $Fm\overline{3}m$ & 3.5 & 3.4 & \cite{day12jssc} & Sr$_{2}$ScNbO$_{6}$ & $P2_{1}/n$ & 3.8 & 3.9 & \cite{barnes06ac}\\
  NaTaO$_{3}$ & $Pcmn$ & 4.4 & 4.0 & \cite{ahtee77ac} & Ba$_{2}$ZnWO$_{6}$ & $Fm\overline{3}m$ & 3.8 & 3.5 & \cite{day12jssc} & Sr$_{2}$YNbO$_{6}$ & $P2_{1}/n$ & 4.0 & 3.9 & \cite{howard05ac}\\
  Sr$_{2}$AlNbO$_{6}$ & $Fm\overline{3}m$ & 3.6 & 3.8 & \cite{woodward94jmr} & Sr$_{2}$GaNbO$_{6}$ & $I4/m$ & 4.1 & 3.8 & \cite{barnes06ac} & Sr$_{2}$YTaO$_{6}$ & $P2_{1}/n$ & 4.6 & 4.7 & \cite{howard05ac}\\
  Ba$_{2}$ScNbO$_{6}$ & $Fm\overline{3}m$ & 3.6 & 3.6 & \cite{barnes06ac} & Sr$_{2}$GaTaO$_{6}$ & $I4/m$ & 4.8 & 4.5 & \cite{barnes06ac} & Sr$_{2}$CaMoO$_{6}$ & $P2_{1}/n$ & 2.8 & 2.7 & \cite{day12jssc}\\
  Ba$_{2}$YNbO$_{6}$ & $Fm\overline{3}m$ & 3.8 & 3.8 & \cite{barnes06ac} & Sr$_{2}$MgMoO$_{6}$ & $I4/m$ & 2.6 & 2.7 & \cite{day12jssc} & Ca$_{3}$WO$_{6}$ & $P2_{1}/n$ & 3.7 & 3.6 & \cite{day12jssc}\\
  Sr$_{2}$AlTaO$_{6}$ & $Fm\overline{3}m$ & 4.3 & 4.6 & \cite{woodward94jmr} & Sr$_{2}$ZnMoO$_{6}$ & $I4/m$ & 2.9 & 2.7 & \cite{day12jssc} & Sr$_{2}$CaWO$_{6}$ & $P2_{1}/n$ & 3.6 & 3.6 & \cite{day12jssc}\\
  Ba$_{2}$YTaO$_{6}$ & $Fm\overline{3}m$ & 4.3 & 4.6 & \cite{zurmuhlen94jap} & Sr$_{2}$MgWO$_{6}$ & $I4/m$ & 3.5 & 3.6 & \cite{day12jssc} & Sr$_{2}$ZnWO$_{6}$ & $P2_{1}/n$ & 3.8 & 3.7 & \cite{day12jssc}\\
  \hline
  Sr$_{2}$AlVO$_{6}$ & $Fm\overline{3}m$ & 1.9 & & & Ba$_{2}$ScVO$_{6}$ & $Fm\overline{3}m$ & 2.1 & & & Sr$_{2}$ZnCrO$_{6}$ & $Fm\overline{3}m$ & 1.7 & & \\
  Sr$_{2}$ScVO$_{6}$ & $Fm\overline{3}m$ & 2.1 & & & Ba$_{2}$GaVO$_{6}$ & $Fm\overline{3}m$ & 2.2 & & & Ba$_{2}$MgCrO$_{6}$ & $Fm\overline{3}m$ & 1.3 & & \\
  Sr$_{2}$GaVO$_{6}$ & $Fm\overline{3}m$ & 2.1 & & & Sr$_{2}$MgCrO$_{6}$ & $Fm\overline{3}m$ & 1.3 & & & Ba$_{2}$CaCrO$_{6}$ & $Fm\overline{3}m$ & 1.3 & & \\
  Ba$_{2}$AlVO$_{6}$ & $Fm\overline{3}m$ & 2.0 & & & Sr$_{2}$CaCrO$_{6}$ & $Fm\overline{3}m$ & 1.3 & & & Ba$_{2}$ZnCrO$_{6}$ & $Fm\overline{3}m$ & 1.5 & & \\
  \hline
  \end{tabular}
  \label{table:testset}
\end{table*}

\begin{figure}
  \begin{center}
    \includegraphics[scale=0.2]{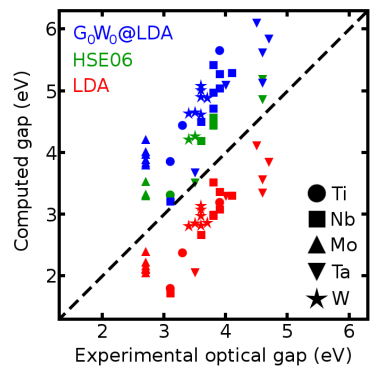}
  \end{center}
  \caption{LDA (red), $G_{0}W_{0}$@LDA (blue), and HSE06 (green) electronic gaps versus experimental optical gaps of compounds synthesized and measured in reference~\cite{eng03jssc}. Compounds are grouped according to the more highly charged B-site cation, the primary component of the conduction band.}
  \label{fig:scatter}
\end{figure}

In Fig.~\ref{fig:scatter}, LDA (red), $G_{0}W_{0}$@LDA (blue), and HSE06 (green) gaps are plotted versus the optical gaps obtained by Eng {\em et al.}~\cite{eng03jssc} Double perovskite compounds are grouped according to the more highly charged B-site cation, whose $d$ states are primary contributors to the conduction band edge. Optical gaps are underestimated by LDA, and significantly overestimated by one-shot $G_{0}W_{0}$@LDA and HSE06. As expected, compounds with smaller, more highly charged cations (e.g., Mo$^{6+}$) tend to have smaller gaps than those with larger, less highly charged cations (e.g., Ta$^{5+}$). An unsatisfactory feature of Fig.~\ref{fig:scatter} is that for all B-site cations, deviations in computed gaps from experiment are large and vary from compound to compound. We find mean average errors of $-0.74 \pm 0.30$, $0.92 \pm 0.36$, and $0.57 \pm 0.25$ eV for LDA, $G_{0}W_{0}$@LDA, and HSE06, respectively. Applying corrections to computed gaps of V$^{5+}$ and Cr$^{6+}$ compounds based on these mean average errors results in estimates consistent with our earlier estimates based on SrTiO$_{3}$. However, as this approach assumes that computed gap errors are randomly distributed, a clearer understanding of the systematic variation of the gap errors within the test set is needed.

\begin{figure*}
  \begin{center}
    \includegraphics[scale=0.2]{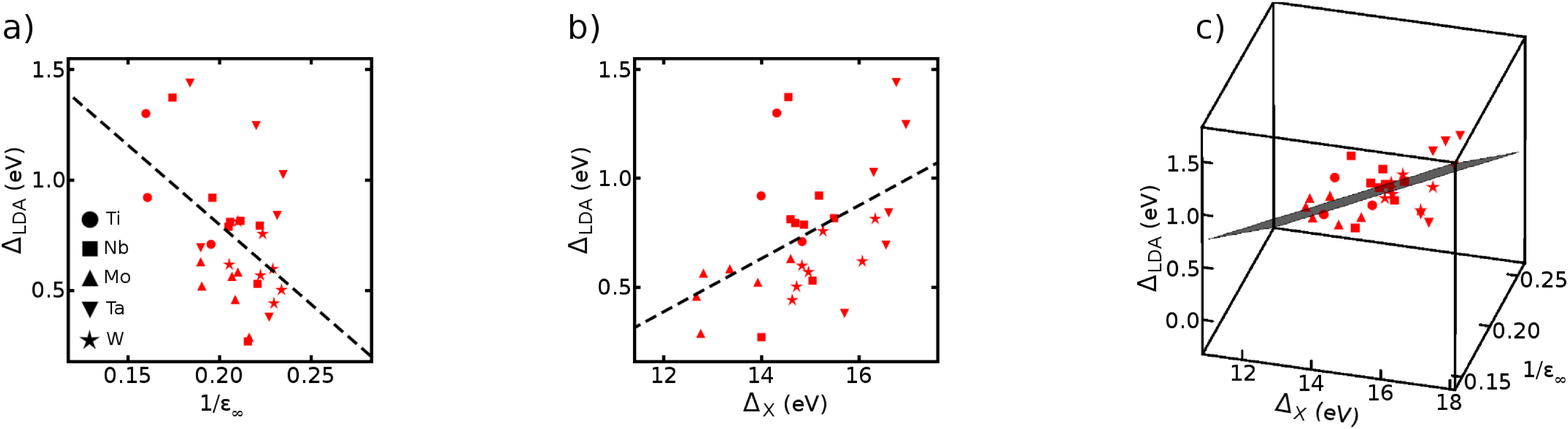}
  \end{center}
  \caption{a) Differences between optical and LDA gaps ($\Delta_\mathrm{LDA}$) versus $\frac{1}{\epsilon_\infty}$ for 30 compounds synthesized and measured in reference~\cite{eng03jssc}. b) $\Delta_\mathrm{LDA}$ versus $\Delta_\mathrm{X}$ (see text). c) $\Delta_\mathrm{LDA}$ versus both $\frac{1}{\epsilon_\infty}$ and $\Delta_\mathrm{X}$. Dashed lines and planes are least-squares fits to the data shown. Compounds are grouped according to the more highly charged B-site cation, the primary component of the conduction band.}
  \label{fig:fits}
\end{figure*}

Past work on traditional semiconductors and insulators has shown that underestimates of experimental gaps by LDA tend to roughly scale as $\frac{1}{\epsilon_\infty}$, where $\epsilon_\infty$ is the electronic component of the static dielectric constant.~\cite{fiorentini95prb} That is, LDA tends to underestimate experimental gaps more severely as $\frac{1}{\epsilon_\infty}$ increases. In contrast, for $d^{0}$ perovskites and double perovskites, Fig.~\ref{fig:fits}a shows that the LDA gap is {\em closer} to experiment as $\frac{1}{\epsilon_\infty}$ increases. The fact that this trend in Fig.~\ref{fig:fits}a runs opposite to established intuition suggests that there is another important factor determining band gap errors in this class of compounds.

Band edge states of $d^{0}$ perovskite-derived compounds are known to vary in their degree of localization, particularly the conduction band minimum (CBM). While cubic ABO$_{3}$ perovskites have CBM states consisting purely of B-site metal cation $d$ character, double perovskite CBM states also include varying amounts of oxygen $2p$ character,~\cite{eng03jssc} leading to a range of localization depending on fractional $d$ and $2p$ content. In Figure~\ref{fig:fits}b, we demonstrate that these differences in band edge localization are partially responsible for the errors in computed band gaps. As a proxy for band edge localization, we compute the difference between the band edge Fock exchange energies ($\Delta_\mathrm{X}$) from the Kohn-Sham LDA wavefunctions associated with the CBM and valence band maximum (VBM). For these compounds, because their VBM exchange energies are all quite similar, a larger value of $\Delta_\mathrm{X}$ indicates that the CBM is derived from more localized $d$ states. Not surprisingly, the data show that a more localized CBM generally results in a more severely underestimated LDA gap. 

In light of Fig.~\ref{fig:fits}b, the counterintuitive trend in Fig.~\ref{fig:fits}a can be rationalized. In this class of $d^{0}$ perovskite-derived compounds, those which have the smallest LDA gaps (and therefore the smallest values of $\frac{1}{\epsilon_\infty}$) are those with the most localized CBM states (and therefore the most severe underestimates of exchange).

In Fig.~\ref{fig:fits}c, we show the result of fitting the LDA gap errors in both $\frac{1}{\epsilon_\infty}$ and $\Delta_\mathrm{X}$ simultaneously, obtaining the following expression to predict the optical gaps of $d^{0}$ perovskite-derived compounds:

\begin{equation}
E_{g}(\mathrm{opt}) = E_{g}(\mathrm{LDA}) + 0.148 \Delta_\mathrm{X} - \frac{8.62 \mathrm{eV}}{\epsilon_\infty} + 0.33 \mathrm{eV} \pm 0.20 \mathrm{eV}.
\label{eqn:est}
\end{equation}

\noindent{Compared to our earlier corrections to computed gaps, this fit reduces the uncertainty of optical gap predictions significantly, to $\pm 0.20$ eV. Estimates based on Eqn.~\ref{eqn:est} of band gaps within the test set are shown in the top section of Table~\ref{table:testset}.}

\subsection{Predicted band gaps of new V$^{5+}$ and Cr$^{6+}$-based compounds}

In Fig.~\ref{fig:hypothetical}c (black) and the bottom section of Table~\ref{table:testset}, we apply these refined band gap estimates to the twelve hypothetical V$^{5+}$ and Cr$^{6+}$ double perovskites discussed earlier. Error bars indicate the uncertainty of the estimates ($\pm 0.20$ eV), as determined by the standard deviations of the fits for known compounds. The results in Fig.~\ref{fig:hypothetical}c indicate that V$^{5+}$ and Cr$^{6+}$ double perovskites with rock-salt B-site ordering would likely have optical gaps spanning significant parts of the solar spectrum (1.7--2.4 eV for V$^{5+}$ and 1.1--1.9 eV for Cr$^{6+}$). While band gaps in this region do not guarantee strong solar absorption, $d^{0}$ perovskite-derived compounds are likely to absorb strongly because their valence and conduction bands consist primarily of oxygen $2p$ and metal $d$ states, respectively.

We note that our band gap estimates from this linear fitting procedure hold for known $d^{0}$ compounds outside the test set.~\footnote{Though there is often a range of optical gaps reported in the literature for a given compound, we compare our estimates to gaps reported based on the onset of optical absorption at room temperature, in order to remain as consistent as possible with our test set.} For cubic perovskite BaZrO$_{3}$, we predict a gap of $4.1 \pm 0.2$ eV, compared to the measured gap of 4.0 eV.~\cite{borja10mse} For ferroelectric perovskite LiNbO$_{3}$, we predict a gap of $4.2 \pm 0.2$ eV, compared to the measured 3.78 eV.~\cite{dhar90jap} The difference between theory and experiment for LiNbO$_{3}$ is not surprising given that zero-temperature calculations tend to exaggerate the amplitude of ferroelectric distortions.~\cite{berger11prl} It is likely that our approach can be used to predict the optical gaps of not only $d^{0}$ perovskites, but of $d^{0}$ transition metal oxides in general. For rutile TiO$_{2}$, we predict a gap of $3.1 \pm 0.2$ eV, compared to the measured value of 3.05 eV.~\cite{cronemeyer52pr}

The fact that $G_{0}W_{0}$@LDA overestimates the optical gaps of our test set of compounds (by an average of 0.92 eV) may stem from several sources. First, $GW$ neglects electron-hole interactions, which are expected to be non-negligible in these systems (on the order of 0.1--0.2 eV in recent calculations of TiO$_{2}$~\cite{kang10prb}). Second, electron-phonon interactions, also neglected in this work, can renormalize the optical gap.~\cite{kang10prb,giustino10prl} Third, it is possible that stronger correlations, inaccessible via the $GW$ approximation, are quantitatively significant in these systems.~\cite{biermann03prl,sun04prl} Finally, a primary assumption of one-shot $G_{0}W_{0}$@LDA is that the LDA wavefunctions are a good approximation to the quasiparticle wavefunctions. Further studies, employing a self-consistent approach to $GW$, and/or more optimal starting points, may improve agreement with experiment.~\cite{jiang10prb} We plan to explore these ideas in future work.

\section{Conclusions}

In summary, we have predicted the band gaps and energetic stability of a class of as-yet unmade A$_{2}$B'BO$_{6}$ double perovskites containing B-site V$^{5+}$ or Cr$^{6+}$. Through a combination of LDA, $G_{0}W_{0}$@LDA, and HSE06 calculations, and analysis of the experimentally determined optical gaps of related $d^{0}$ perovskite-derived compounds, we have determined that a number of these hypothetical V$^{5+}$ and Cr$^{6+}$ compounds are likely to be metastable at room temperature with respect to distortion and changes in B-site ordering, and to have optical gaps spanning significant portions of the solar spectrum. If compounds such as these could be synthesized, whether in bulk or epitaxially, they would likely be of interest in light harvesting and energy conversion applications.

\section{Acknowledgments}

We thank C.~J. Fennie and D.~G. Schlom for valuable discussions. This material is based upon work supported as part of the Energy Materials Center at Cornell (EMC$^{2}$), an Energy Frontier Research Center funded by the U.S.~Department of Energy, Office of Science, Office of Basic Energy Sciences under Award Number DE-SC0001086. Work at the Molecular Foundry was supported by the Office of Science, Office of Basic Energy Sciences of the U.S.~Department of Energy under Contract Number DE-AC02-05CH11231.

\section{Appendix: Metastability of Predicted Compounds}

\begin{figure}
  \begin{center}
    \includegraphics[scale=0.2]{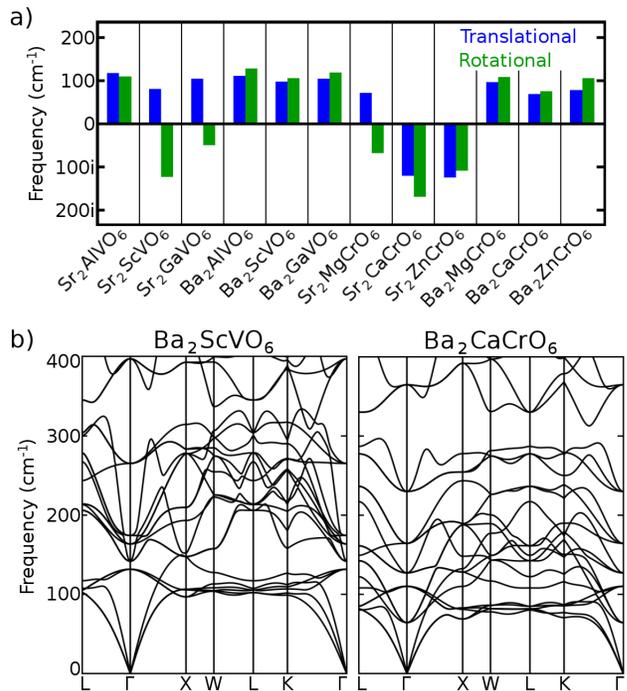}
  \end{center}
  \caption{a) The most unstable phonon frequencies at the zone center of a $2\times2\times2$ perovskite unit cell for twelve hypothetical V$^{5+}$ and Cr$^{6+}$ double perovskites with rock-salt B-site ordering, with phonon modes in which B-site cations move (``translational'') shown separately from others (``rotational''). b) Computed phonon dispersion relations of Ba$_{2}$ScVO$_{6}$ (left) and Ba$_{2}$CaCrO$_{6}$ (right) with undistorted rock-salt B-site ordering. Real frequencies indicate stability with respect to distortion.}
  \label{fig:dispersion}
\end{figure}

In what follows, we comment on the stability of V$^{5+}$ and Cr$^{6+}$ double perovskites studied in this work. For the rock-salt B-site ordering of our twelve hypothetical V$^{5+}$ and Cr$^{6+}$ compounds, structures are relaxed in DFT-LDA within $Fm\overline{3}m$ space group symmetry. All elastic constants (i.e., matrix elements of the stiffness tensor) are found to be non-negative. The most unstable phonon frequencies at the zone center of a $2\times2\times2$ perovskite unit cell are shown in Fig.~\ref{fig:dispersion}a, separated between phonon modes that are associated with the movement of B-site atoms (``translational'') and those that are not (``rotational''). Seven of the twelve compounds have only real phonon frequencies, indicating stability with respect to structural distortions. The five compounds with imaginary zone-center phonon frequencies exhibit soft octahedral rotation modes of a magnitude similar to SrTiO$_{3}$ ($91i$ cm$^{-1}$ in LDA), where these modes are suppressed well below room temperature.~\cite{fleury68prl,muller68prl} Two compounds (Sr$_{2}$CaCrO$_{6}$ and Sr$_{2}$ZnCrO$_{6}$) have unstable distortions that break inversion symmetry; we note that ferroelectric oxides with small band gaps may be of potential interest for their ability to separate electrons and holes.~\cite{bennett08jacs,xu11epl} Fig.~\ref{fig:dispersion}b shows full phonon dispersion relations for Ba$_{2}$ScVO$_{6}$ and Ba$_{2}$CaCrO$_{6}$, which indicate stability throughout the Brillouin zone. Due to their structural and chemical similarity, we expect qualitatively similar behavior for the other hypothetical compounds. These results suggest that most of these hypothetical cubic double perovskites with rock-salt B-site ordering are likely metastable at room temperature.

\begin{figure}
  \begin{center}
    \includegraphics[scale=0.2]{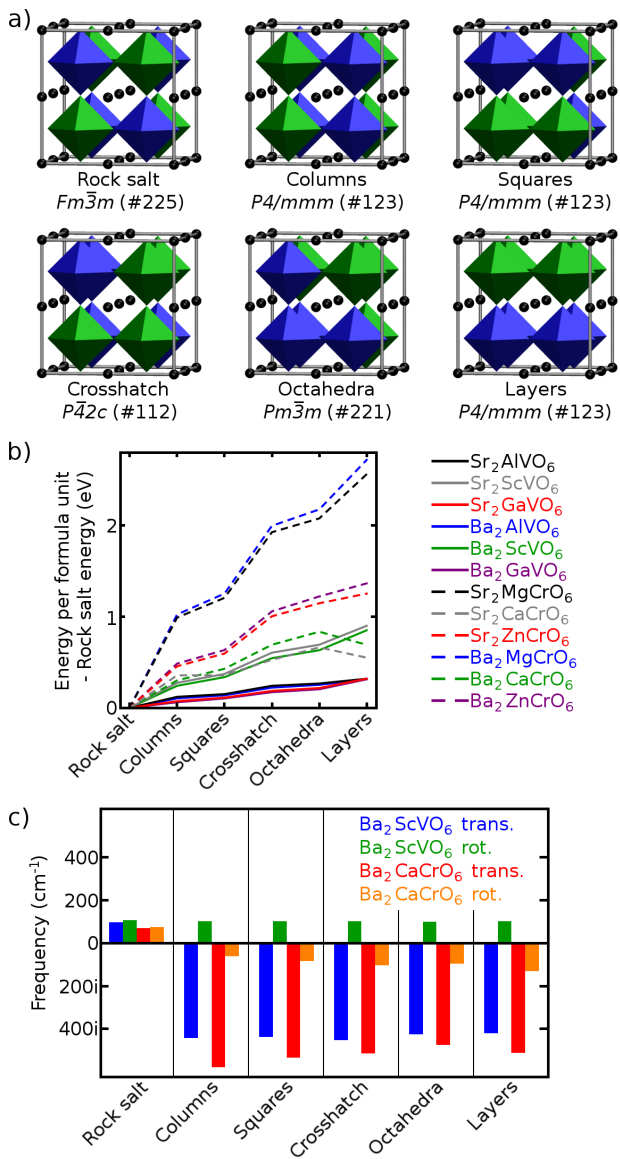}
  \end{center}
  \caption{a) The six distinct B-site orderings of A$_{2}$B'BO$_{6}$ double perovskites in a $2\times2\times2$ perovskite unit cell. b) Total energies per formula unit, relative to the rock-salt B-site ordering of each compound and in the absence of symmetry-breaking distortions. c) The most unstable phonon frequencies at the zone center of a $2\times2\times2$ perovskite unit cell for the distinct B-site orderings of Ba$_{2}$ScVO$_{6}$ and Ba$_{2}$CaCrO$_{6}$, with phonon modes in which B-site cations move (``translational'') shown separately from others (``rotational'').}
  \label{fig:ordering}
\end{figure}

To assess whether rock-salt ordering is indeed the most stable arrangement of B-site atoms, the total energy associated with this ordering was compared to those of the five other distinct B-site orderings possible in a $2\times2\times2$ perovskite unit cell (Fig.~\ref{fig:ordering}a). In the absence of symmetry-breaking distortions, rock-salt ordering has the lowest total energy for all twelve combinations of elements (Fig.~\ref{fig:ordering}b).

\begin{figure}
  \begin{center}
    \includegraphics[scale=0.2]{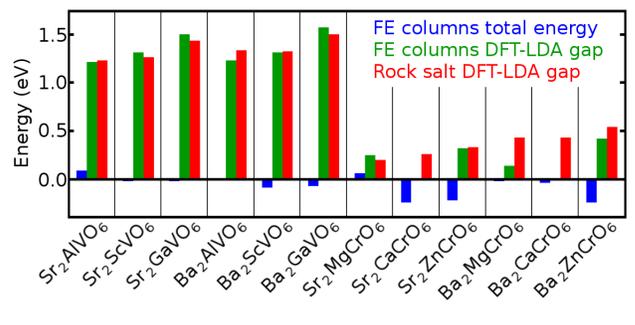}
  \end{center}
  \caption{The DFT-LDA total energy per formula unit of the ferroelectric (FE) columnar ordering of twelve hypothetical V$^{5+}$ and Cr$^{6+}$ double perovskites relative to rock-salt ordering (blue), and DFT-LDA gaps of both geometries (green, red).}
  \label{fig:estimate}
\end{figure}

However, there is still the possibility of distortions (translational or rotational) changing the relative energetics of these orderings. To explore this possibility, Ba$_{2}$ScVO$_{6}$ and Ba$_{2}$CaCrO$_{6}$ were subjected to more thorough stability analysis. Their most unstable phonon frequencies at the zone center of a $2\times2\times2$ perovskite unit cell are shown in Fig.~\ref{fig:ordering}c for all six B-site orderings in the absence of distortion. In all cases, elastic constants are non-negative. Based on these results, we expect octahedral rotations to have little effect on the relative energetics, as they are equally present in all orderings (and possibly suppressed at room temperature as in SrTiO$_{3}$). Structures were then relaxed along all unstable translational phonon modes along high-symmetry axes. For both Ba$_{2}$ScVO$_{6}$ and Ba$_{2}$CaCrO$_{6}$, the only structures whose energies compete favorably with those of rock-salt ordering are distorted variants of columnar ordering (Fig.~\ref{fig:ordering}a, top middle). In both cases, the structure with the lowest total energy is the columnar structure with ferroelectric polarization along the direction of the columns. Both compounds in this geometry have non-negative elastic constants and real phonon frequencies, meaning they are not expected to distort further. In Fig.~\ref{fig:estimate}, the rock-salt and polarized columnar B-site orderings of all twelve hypothetical compounds are compared in both total energy (blue) and DFT-LDA band gap (green, red). For three of the twelve compounds (Sr$_{2}$AlVO$_{6}$, Ba$_{2}$AlVO$_{6}$, and Sr$_{2}$MgCrO$_{6}$), rock-salt B-site ordering is the most stable structure. For the remaining nine, a few points are worth noting. First, it is possible that internal distortions would be sufficiently suppressed at room temperature as to favor rock-salt ordering. Second, the computed DFT-LDA gaps of the two competing structures are quite similar in most cases, suggesting that even if present, this change in B-site ordering and distortion would do little to change measured band gaps. Third, even if it is not the lowest-energy structure, rock-salt B-site ordering could in principle be achieved via epitaxial growth in the $[111]$ direction.

\bibliographystyle{prsty}
\bibliography{VCr.bib}


\end{document}